\documentclass[pre,twocolumn,aps,showpacs]{revtex4}
\newcommand{\bec}[1]{\mbox{\boldmath $ #1$}}
\usepackage{graphicx}
\begin{document}
\title{Electromotive force and large-scale magnetic dynamo
in a turbulent flow with a mean shear}
\author{Igor Rogachevskii}
\email{gary@menix.bgu.ac.il}
\homepage{http://www.bgu.ac.il/~gary}
\author{Nathan Kleeorin}
\email{nat@menix.bgu.ac.il}
\affiliation{Department of Mechanical
Engineering, The Ben-Gurion
University of the Negev, \\
POB 653, Beer-Sheva 84105, Israel}
\date{\today}
\begin{abstract}
An effect of sheared large-scale motions on a mean electromotive
force in a nonrotating turbulent flow of a conducting fluid is
studied. It is demonstrated that in a homogeneous divergence-free
turbulent flow the $\alpha$-effect does not exist, however a mean
magnetic field can be generated even in a nonrotating turbulence
with an imposed mean velocity shear due to a new ''shear-current"
effect. A mean velocity shear results in an anisotropy of
turbulent magnetic diffusion. A contribution to the electromotive
force related with the symmetric parts of the gradient tensor of
the mean magnetic field (the $\kappa$--effect) is found in a
nonrotating turbulent flows with a mean shear. The
$\kappa$--effect and turbulent magnetic diffusion reduce the
growth rate of the mean magnetic field. It is shown that a mean
magnetic field can be generated when the exponent of the energy
spectrum of the background turbulence (without the mean velocity
shear) is less than 2. The ''shear-current" effect was studied
using two different methods: the $\tau$--approximation (the Orszag
third-order closure procedure) and the stochastic calculus (the
path integral representation of the solution of the induction
equation, Feynman-Kac formula and Cameron-Martin-Girsanov
theorem). Astrophysical applications of the obtained results are
discussed.
\end{abstract}


\maketitle

\section{Introduction}

   Generation of magnetic fields by a turbulent flow of a conducting
fluid is a fundamental problem which has a number of applications
in solar physics and astrophysics, geophysics and planetary
physics (see, {\em e.g.,} \cite{M78,P79,KR80,ZRS83,RSS88}). It is
known that small-scale magnetic fluctuations with a zero mean
magnetic field can be generated in a homogeneous nonhelical and
nonrotating turbulence by a stretch-twist-fold mechanism (see,
{\em e.g.,} \cite{ZMR88,ZRS90,CG95,BR96,RK97,KRS02}). On the other
hand, in a homogeneous divergence-free turbulent flow the helicity
and the $\alpha$-effect vanish.

However, the mean magnetic field can be generated in a rotating
homogeneous turbulent flow due to the combined action of the ${\bf
\Omega \times J}$--effect and a nonuniform (differential) rotation
\cite{R69,R72,MP82,R86,RKR02}, where ${\bf \Omega}$ is the angular
velocity and ${\bf J}$ is the mean electric current. The evolution
of the mean magnetic field $ \bar{\bf B} $ is determined by
equation
\begin{eqnarray}
{\partial \bar{\bf B} \over \partial t} = \bec{\nabla} \times
(\bar{\bf U} {\bf \times} \bar{\bf B} + \bec{\cal E} - D_m
\bec{\nabla} {\bf \times} \bar{\bf B}) \;, \label{B3}
\end{eqnarray}
where $\bar{\bf U}$ is the mean velocity, $D_m$ is the magnetic
diffusion due to the electrical conductivity of fluid, $ \bec{\cal
E} = \langle {\bf u} \times {\bf b} \rangle $ is the mean
electromotive force. The general form of the mean electromotive
force was suggested in \cite{R80} using the symmetry arguments:
\begin{eqnarray}
{\cal E}_{i} &=& {\alpha}_{ij} \bar B_{j} - {\beta}_{ij}
(\bec{\nabla} {\bf \times} \bar {\bf B})_{j} + ({\bf V}^{\rm
(eff)} {\bf \times} \bar {\bf B})_{i}
\nonumber\\
&& - [\bec{\delta} {\bf \times} (\bec{\nabla} {\bf \times}
\bar{\bf B})]_i - {\kappa}_{ijk} ({\partial \hat B})_{jk} \;,
\label{C4}
\end{eqnarray}
where $({\partial \hat B})_{ij} = (\nabla_i \bar B_{j} + \nabla_j
\bar B_{i}) / 2 ,$ $ {\bf u} $ and $ {\bf b} $ are fluctuations of
the velocity and magnetic field, respectively, angular brackets
denote averaging over an ensemble of turbulent fluctuations, the
tensors $ \alpha_{ij} $ and $ \beta_{ij} $ describe the $ \alpha
$-effect and turbulent magnetic diffusion, respectively, $ {\bf
V}^{\rm eff} $ is the effective diamagnetic (or paramagnetic)
velocity, $ \kappa_{ijk} $ and $ \bec{\delta} $ describe a
nontrivial behavior of the mean magnetic field in an anisotropic
turbulence. The ${\bf \Omega \times J}$--effect, e.g., is
associated with the $ \bec{\delta} $-term in the mean
electromotive force.

In the present paper we suggested a new mechanism of generation of
a mean magnetic field by a {\em nonrotating homogeneous} and
nonhelical turbulence with an imposed mean velocity shear. This
mechanism is associated with a ''shear-current" effect determined
by the $ \bec{\delta} $-term in the mean electromotive force. On
the other hand, the turbulent magnetic diffusion and the
$\kappa$--effect can reduce the growth rate of the mean magnetic
field. The $\kappa$--effect arises in an anisotropic turbulence
caused by the mean velocity shear. Our analysis of the mean-field
magnetic dynamo showed that the generation of a mean magnetic
field can occur when $q<2 ,$ where $q$ is the exponent of the
energy spectrum of the background turbulence (without a mean
velocity shear). In particular, in Kolmogorov background
turbulence with $q=5/3$ a mean magnetic field can be generated.
The ''shear-current" effect was studied using two different
methods: the $\tau$--approximation (the Orszag third-order closure
procedure \cite{O70}, see Section IV) and the stochastic calculus
(the path integral representation of the solution of the induction
equation, Feynman-Kac formula and Cameron-Martin-Girsanov theorem,
see Appendixes A and B). We also calculated the mean electromotive
force for an arbitrary weakly inhomogeneous turbulence with an
imposed mean velocity shear. The inhomogeneity of turbulence and
mean velocity shear cause the $\alpha$--effect and the effective
drift velocity of the mean magnetic field.

The $\bec{\delta}$--term in the electromotive force which is
responsible for the ''shear-current" effect was also independently
found in \cite{RS02} in a problem of a screw dynamo using the
modified second-order correlation approximation. Note also that
for homogeneous and nonhelical flows another mechanism for the
magnetic dynamo associated with a ''negative turbulent magnetic
diffusivity" was recently discussed in \cite{LN99,ZP01,U02}.

This paper is organized as follows. In Section II the general form
of the mean electromotive force which includes the shear-current
effect is obtained using simple symmetry reasoning, and the
mechanism for the shear-current effect is also discussed. In
Section III the governing equations for turbulent velocity and
magnetic fields are formulated which then are used for study an
effect of a mean velocity shear on a turbulence (Section IV) and
on a cross-helicity (Section V). This allows us to determine the
mean electromotive force in a turbulent flow of a conducting fluid
with an imposed mean velocity shear (Section VI). The implications
of the results for the mean electromotive force to the mean-field
magnetic dynamo in a nonrotating homogeneous turbulence are
performed in Section VII. Conclusions and astrophysical
applications of the obtained results are discussed in Section
VIII. In Appendixes A and B the ''shear-current" effect is studied
using another approach, i.e., stochastic calculus. In Appendix C
we derived identities used for the derivation of equations for the
second moment of the velocity field and the cross-helicity tensor.

\section{The qualitative description}

The mean electromotive force can be written in the form:
\begin{eqnarray}
{\cal E}_{i} =  a_{ij} \bar B_{j} + b_{ijk} \bar B_{j,k} +
O(\nabla^2 \bar B_{i}) \; .
\label{K1}
\end{eqnarray}
Following to \cite{R80} we use an identity $ \bar B_{j,k} =
(\partial \hat B)_{jk} - \varepsilon_{jkl} (\bec{\nabla} {\bf
\times} \bar {\bf B})_{l} / 2 $ which allows us to rewrite
Eq.~(\ref{K1}) for the mean electromotive force in the form of
Eq.~(\ref{C4}), where $\varepsilon_{ijk}$ is the fully
antisymmetric Levi-Civita tensor, and
\begin{eqnarray}
\alpha_{ij} &=& (a_{ij} + a_{ji}) / 2 \;,
\label{C5} \\
\beta_{ij} &=& (\varepsilon_{ikp} b_{jkp} + \varepsilon_{jkp}
b_{ikp}) / 4 \;,
\label{C6} \\
V^{\rm (eff)}_{k} &=& \varepsilon_{kji} a_{ij} / 2 \;,
\label{C7}  \\
\delta_{i} &=& (b_{jji} - b_{jij}) / 4 \;,
\label{C8}  \\
\kappa_{ijk} &=& - (b_{ijk} + b_{ikj}) / 2 \; .
\label{C9}
\end{eqnarray}
In a homogeneous and nonhelical turbulence the tensor $a_{ij}$
vanishes, which implies that $\alpha_{ij} = 0 $ and $V^{\rm
(eff)}_{k} = 0 .$ Below we consider this case.

The general form of the mean electromotive force in a turbulent
flow with a mean velocity shear can be obtained even from simple
symmetry reasoning. Indeed, the electromotive force $\bec{\cal E}$
is a true vector, whereas the mean magnetic field $\bar{\bf B}$ is
a pseudo-vector. Therefore, the tensor $b_{ijk}$ is a
pseudo-tensor (see Eq.~(\ref{K1})). For homogeneous, isotropic and
nonhelical turbulence the tensor $b_{ijk} = \beta_{_{T}}
\varepsilon_{ijk} ,$ where $\beta_{_{T}}$ is the turbulent
magnetic diffusion coefficient. In a turbulent flow with an
imposed mean velocity shear, the tensor $b_{ijk}$ depends on the
true tensor $\nabla_j \bar U_{i} .$ Note that the tensor $
\nabla_j \bar U_{i}  $ can be written as a sum of the symmetric
and antisymmetric parts, i.e., $ \nabla_j \bar U_{i} = (\delta
\hat U)_{ij} - (1/2) \varepsilon_{ijk} \, \bar W_{k} ,$ where $
(\delta \hat U)_{ij} = (\nabla_i \bar U_{j} + \nabla_j \bar U_{i})
/ 2 $ is the true tensor and $\bar{\bf W} = \bec{\nabla} {\bf
\times} \bar {\bf U}$ is the mean vorticity (pseudo-vector).
Hereafter we take into account the effect which is linear in $
\nabla_j    \bar U_{i} .$ Therefore, the pseudo-tensor $b_{ijk}$
has the following general form
\begin{eqnarray}
b_{ijk} &=& \beta_{_{T}} \varepsilon_{ijk} + l_0^2 \, [a_1 \,
\varepsilon_{ijm} (\delta \hat U)_{mk} + a_2 \, \varepsilon_{ikm}
(\delta \hat U)_{mj}
\nonumber\\
&& + a_3 \, \varepsilon_{jkm} (\delta \hat U)_{mi} + a_4 \,
\delta_{ij} \bar W_{k} + a_5 \, \delta_{ik} \bar W_{j}] \;,
\label{K2}
\end{eqnarray}
where $ a_k $ are the unknown coefficients, $l_0$ is the maximum
scale of turbulent motions, and the term $\propto \delta_{jk} \bar
W_{i}$ vanishes since $ \bec{\nabla} \cdot \bar{\bf B} = 0 $ (see
Eq.~(\ref{K1})). Using Eqs.~(\ref{C5})-(\ref{K2}) we determine the
turbulent coefficients defining the mean electromotive force for a
homogeneous and nonhelical turbulence:
\begin{eqnarray}
\beta_{ij} &=& \beta_{_{T}} \, \delta_{ij} - 2 \, \beta_{0} \,
l_0^2 \, (\partial \hat U)_{ij} \;,
\label{K3}\\
\bec{\delta} &=& l_0^2 \, \delta_0 \, \bar{\bf W} \;,
\label{K4}\\
\kappa_{ijk} &=& l_0^2 \, (\kappa_{1} \, \delta_{ij} \, \bar W_k +
\kappa_{2} \, \varepsilon_{ijm} \, (\partial \hat U)_{mk})  \;,
\label{K5}
\end{eqnarray}
where
\begin{eqnarray}
\beta_{0} &=& (a_1 - a_2 - 2 a_3) / 4 \;, \quad \delta_0 = (a_4 -
a_5) / 2  \;,
\label{K20}\\
\kappa_{1} &=& - (a_4 + a_5) \;, \quad \kappa_{2} = - (a_1 + a_2)
\;,
\label{K21}
\end{eqnarray}
and $\beta_{_{T}} = u_0 l_0 / 3$ is the coefficient of isotropic
part of turbulent magnetic diffusion, while the second term in
Eq.~(\ref{K3}) determines anisotropic part of turbulent magnetic
diffusion caused by the mean velocity shear. Here $u_0$ is the
characteristic turbulent velocity in the maximum scale of
turbulent motions. The $\kappa$--effect (determined by the tensor
$\kappa_{ijk})$ describes a contribution to the electromotive
force related with the symmetric parts of the gradient tensor of
the mean magnetic field and arises in an anisotropic turbulence
caused by the mean velocity shear. Since the tensor $\kappa_{ijk}$
is multiplied by the symmetric tensor $(\partial \hat B)_{jk} $ in
the the mean electromotive force, this allows us to rewrite the
tensor $\kappa_{ijk}$ determined by Eq.~(\ref{K5}) in a more
simple but not in a symmetric form. We will show in this paper
that the $\bec{\delta}$--term in Eqs.~(\ref{C4}) and~(\ref{K4})
for the mean electromotive force describes the ''shear-current"
effect which can cause the mean-field magnetic dynamo in a
homogeneous nonrotating turbulence with an imposed mean velocity
shear.

Consider a homogeneous divergence-free turbulence with a mean
velocity shear, e.g., $ \bar{\bf U} = (0, Sx, 0)$ and $ \bar{\bf
W} = (0,0,S) .$ The mean magnetic field is determined by equation:
\begin{eqnarray}
{\partial \bar{\bf B} \over \partial t} &=& \bec{\nabla} {\bf
\times} [\bar{\bf U} {\bf \times} \bar{\bf B} - \bec{\hat \beta}
(\bec{\nabla} {\bf \times} \bar{\bf B})
\nonumber\\
&& - \bec{\delta} {\bf \times} (\bec{\nabla} {\bf \times} \bar{\bf
B}) - \bec{\hat \kappa}  ({\partial \hat B})] \;, \label{E1}
\end{eqnarray}
where $ \bec{\hat \beta} \equiv \beta_{ij} $ and $ \bec{\hat
\kappa} \equiv \kappa_{ijk} .$ Now for simplicity we use the mean
magnetic field in the form $ \bar{\bf B} = (\bar B_x(z), \bar
B_y(z), 0) .$ Then Eq.~(\ref{E1}) reads
\begin{eqnarray}
{\partial \bar B_x \over \partial t} &=& - S \, l_0^2 \, \sigma_0
\, \bar B''_y + \beta_{_{T}} \, \bar B''_x  \;,
\label{E2}\\
{\partial \bar B_y \over \partial t} &=& S \, \bar B_x +
\beta_{_{T}} \, \bar B''_y  \;,
\label{E3}
\end{eqnarray}
where
\begin{eqnarray}
\sigma_0 = \delta_0 - \beta_0 - \kappa_1 / 2 - \kappa_2 / 4 \;,
\label{K22}
\end{eqnarray}
and $ \bar B'' = \partial^2 \bar B / \partial z^2 .$  In
Eq.~(\ref{E3}) we took into account that $ l_0^2 \bar B''_x \ll
\bar B_x ,$ i.e., the characteristic spatial scale $ L_B $ of the
mean magnetic field variations is much larger than the maximum
scale of turbulent motions $ l_0 .$ This assumption corresponds to
the mean-field approach. The first term $ (\propto S \bar B_x) $
in RHS of Eq.~(\ref{E3}) plays a role of the differential
rotation. Indeed, $ \bec{\nabla} {\bf \times} (\bar{\bf U} {\bf
\times} \bar{\bf B}) = (\bar{\bf B} \cdot \bec{\nabla}) \bar{\bf
U} - (\bar{\bf U} \cdot \bec{\nabla}) \bar{\bf B} = S \bar B_x
{\bf e}_y ,$ and for the chosen configuration of the mean magnetic
field, $(\bar{\bf U} \cdot \bec{\nabla}) \bar{\bf B} = 0 .$

A solution of Eqs.~(\ref{E2}) and~(\ref{E3}) we seek for in the
form $ \propto \exp(\gamma t + i K z) ,$ where $\gamma$ is given
by
\begin{eqnarray}
\gamma = S l_0 K \sqrt{\sigma_0} - \beta_{_{T}} K^2 \;,
\label{K6}
\end{eqnarray}
where $\sigma_0 = (a_2 + a_3 + 2 a_4) / 2 .$ The magnetic dynamo
instability can be excited when $\sigma_0 > 0 .$ In this paper we
will find unknown coefficients $ a_k ,$ which will allow us to
determine the conditions for the generation of the mean magnetic
field due to the magnetic dynamo instability caused by the
''shear-current" effect.

In order to elucidate the physics of the ''shear-current" effect,
let us compare the $\alpha$-term in the electromotive force which
is responsible for the generation of the mean magnetic field,
i.e.,
\begin{eqnarray}
{\cal E}^\alpha_i \equiv \alpha \bar B_i \propto - ({\bf \Omega}
\cdot {\bf \Lambda}) \bar B_i \;
\label{K7}
\end{eqnarray}
(see, e.g., \cite{KR80,RKR02}), with the $\bec{\delta}$-term in
the electromotive force caused by the ''shear-current" effect,
i.e.,
\begin{eqnarray}
{\cal E}^\delta_i \equiv - (\bec{\delta} {\bf \times}
(\bec{\nabla} {\bf \times} \bar{\bf B}))_i \propto - (\bar{\bf W}
\cdot \bec{\nabla}) \bar B_i \;,
\label{K8}
\end{eqnarray}
where $ {\bf \Lambda} = \bec{\nabla} \langle {\bf u}^2 \rangle /
\langle {\bf u}^2 \rangle $ determines the inhomogeneity of
turbulence. Here for simplicity we considered an isotropic
$\alpha$-tensor, i.e., $ \alpha_{ij} = \alpha \delta_{ij} .$ There
is an analogy between the $\alpha$-term and the
$\bec{\delta}$-term in the electromotive force. In particular, the
mean vorticity $\bar{\bf W}$ plays a role of rotation ${\bf
\Omega}$ and an inhomogeneity of the mean magnetic field plays a
role of the inhomogeneity of turbulence in the $\alpha {\bf
\Omega} $-dynamo (see below). During the generation of the mean
magnetic field in both cases, the mean electric current along the
original mean magnetic field arises. The $\alpha$-effect is
related with the hydrodynamic helicity $ \propto ({\bf \Omega}
\cdot {\bf \Lambda}) $ in an inhomogeneous turbulence. The
deformation of the magnetic field lines is caused by upward and
downward rotating turbulent eddies. Since the turbulence is
inhomogeneous (which breaks a symmetry between the upward and
downward eddies), their total effect on the mean magnetic field
does not vanish and it creates the mean electric current along the
original mean magnetic field.

In a turbulent flow with an imposed mean velocity shear, the
inhomogeneity of the original mean magnetic field breaks a
symmetry between the influence of upward and downward turbulent
eddies on the mean magnetic field. The deformation of the magnetic
field lines is caused by upward and downward turbulent eddies
which causes the mean electric current along the mean magnetic
field and produces the magnetic dynamo. The magnetic dynamo
instability due to the ''shear-current" effect is determined by a
system of Eqs.~(\ref{E2}) and~(\ref{E3}), and there is a coupling
between the components of the mean magnetic field. In particular,
the field $ \bar B_y $ generates the field $ \bar B_x $ due to the
''shear-current" effect (see Eq.~(\ref{E2})). This is similar to
the $\alpha$ effect. On the other hand, the field $ \bar B_x $
generates the field $ \bar B_y $ due to the pure shear effect (see
Eq.~(\ref{E3})), like the differential rotation in $\alpha
\Omega$-dynamo.

In the next sections we will describe the above magnetic dynamo
effect quantitatively using two different methods: the
$\tau$--approximation (the Orszag third-order closure procedure
\cite{O70}) and the stochastic calculus (the path integral
representation of the solution of the induction equation,
Feynman-Kac formula and Cameron-Martin-Girsanov theorem).

\section{The governing equations}

Our goal is to study an effect of sheared large-scale motions on a
mean electromotive force in a nonrotating turbulent flows of a
conducting fluid. The momentum equation for the fluid velocity
${\bf v}$ and the induction equation for the magnetic field ${\bf
h}$ read
\begin{eqnarray}
\biggl({\partial \over \partial t} + {\bf v} \cdot
\bec{\nabla}\biggr){\bf v} &=& - {\bec{\nabla} P \over \rho} +
{\bf F}_m({\bf h}) + \nu \Delta {\bf v}  + {\bf F}^{\rm (st)} ,
\nonumber\\
\label{B1} \\
\biggl({\partial \over \partial t} + {\bf v} \cdot
\bec{\nabla}\biggr) {\bf h} &=& ({\bf h} \cdot \bec{\nabla}) {\bf
v} + D_m \Delta {\bf h} \;, \label{B2}
\end{eqnarray}
where $\bec{\nabla} \cdot {\bf v} = 0 $ and $ \bec{\nabla} \cdot
{\bf h} = 0 ,$ $\, \nu$ is the kinematic viscosity, ${\bf
F}_m({\bf h}) = - (1 / \mu \rho) [{\bf h} {\bf \times}
(\bec{\nabla} {\bf \times} {\bf h})] $ is the magnetic force, $
\mu $ is the magnetic permeability of the fluid, $ {\bf F}^{\rm
(st)} $ is the random external stirring force, $\, P$ and $\rho$
are the pressure and density of fluid, respectively.

We will use a mean field approach whereby the velocity, pressure
and magnetic field are separated into the mean and fluctuating
parts: $ {\bf v} = \bar{\bf U} + {\bf u} ,$ $ \, P = \bar P + p ,$
and $ {\bf h} = \bar{\bf B} + {\bf b} ,$ the fluctuating parts
have zero mean values, and $ \bar{\bf U} = \langle {\bf v} \rangle
,$ $ \, \bar P = \langle P \rangle ,$ $\, \bar{\bf B} = \langle
{\bf h} \rangle .$ Averaging Eqs.~(\ref{B1}) and (\ref{B2}) over
an ensemble of fluctuations we obtain the mean-field equations. In
particular, the evolution of the mean magnetic field $ \bar{\bf B}
$ is determined by Eq.~(\ref{B3}), where $ \bec{\cal E} = \langle
{\bf u} \times {\bf b} \rangle $ is the mean electromotive force.
To determine the mean electromotive force we use equations for
fluctuations $ {\bf u}(t,{\bf r}) $ and $ {\bf b}(t,{\bf r}) $
which are obtained by subtracting equations for the mean fields
from the corresponding equations (\ref{B1}) and (\ref{B2}) for the
total fields:
\begin{eqnarray}
{\partial {\bf u} \over \partial t} &=& - (\bar{\bf U} \cdot
\bec{\nabla}) {\bf u} - ({\bf u} \cdot \bec{\nabla}) \bar{\bf U} -
{\bec{\nabla} p \over \rho} + {\bf F}_m({\bf b},\bar{\bf B})
\nonumber\\
& & + {\bf F}^{\rm (st)} + {\bf U}_{N} \;,
\label{B5} \\
{\partial {\bf b} \over \partial t} &=& - (\bar{\bf U} \cdot
\bec{\nabla}) {\bf b} + ({\bf b} \cdot \bec{\nabla}) \bar{\bf U} -
({\bf u} \cdot \bec{\nabla}) \bar{\bf B}
\nonumber\\
& & + (\bar{\bf B} \cdot \bec{\nabla}) {\bf u} + {\bf B}_{N} \;,
\label{B6}
\end{eqnarray}
where
\begin{eqnarray*}
{\bf F}_m({\bf b},\bar{\bf B}) &=& - {1 \over \mu \rho}[{\bf b}
{\bf \times} (\bec{\nabla} {\bf \times} \bar{\bf B}) + \bar{\bf B}
{\bf \times} (\bec{\nabla} {\bf \times} {\bf b})] \;,
\\
U_{N} &=& \langle ({\bf u} \cdot \bec{\nabla}) {\bf u} \rangle -
({\bf u} \cdot \bec{\nabla}) {\bf u} + [\langle {\bf b} {\bf
\times} (\bec{\nabla} {\bf \times} {\bf b}) \rangle
\\
& & - {\bf b} {\bf \times} (\bec{\nabla} {\bf \times} {\bf b})] /
(\mu \rho) + \nu \Delta {\bf u} \;,
\\
B_{N} &=& \bec{\nabla} {\bf \times} ({\bf u} {\bf \times} {\bf b}
- \langle {\bf u} {\bf \times} {\bf b} \rangle) + D_m \Delta {\bf
b}  \; .
\end{eqnarray*}
We consider a turbulent flow with large hydrodynamic $ ({\rm Re} =
l_{0} u_{0} / \nu \gg 1) $ and magnetic $ ({\rm Rm} = l_{0} u_{0}
/ D_m \gg 1)$ Reynolds numbers, where $ u_{0} $ is the
characteristic velocity in the maximum scale $ l_{0} $ of
turbulent motions. In the next sections we will use
Eqs.~(\ref{B5}) and~(\ref{B6}) to study an effect of a mean
velocity shear on a turbulence (Section IV) and on a
cross-helicity (Section V) in order to determine the mean
electromotive force.

\section{Effect of a mean velocity shear on a turbulence}

In this section we study an effect of a mean velocity shear on a
turbulence using Eq.~(\ref{B5}). We neglect an effect of the mean
magnetic field on the turbulence, i.e., we neglect the magnetic
force ${\bf F}_m({\bf b},\bar{\bf B})$ in Eq.~(\ref{B5}). This is
valid when $\bar B^2 / \mu \ll \rho \langle u^2 \rangle / 2 ,$
i.e., we do not consider the quenching effects (see, e.g.,
\cite{GD94,FB99,K99,RK01,KMRS02,BB02}). We use a two-scale
approach, i.e., a correlation function is written as follows
\begin{eqnarray*}
\langle u_i({\bf x}) u_j ({\bf  y}) \rangle &=& \int \langle u_i
({\bf k}_1) u_j ({\bf k}_2) \rangle  \exp[i({\bf  k}_1 {\bf \cdot}
{\bf x} \\
&& + {\bf k}_2 {\bf \cdot} {\bf y})] \,d{\bf k}_1 \, d{\bf k}_2
\\
&=& \int f_{ij}({\bf k, R}) \exp(i {\bf k} {\bf \cdot} {\bf r})
\,d {\bf k} \;,
\\
f_{ij}({\bf k, R}) &=& \int \langle u_i ({\bf k} + {\bf  K} / 2)
u_j(-{\bf k} + {\bf  K} / 2) \rangle
\\
&& \times \exp(i {\bf K} {\bf \cdot} {\bf R}) \,d {\bf K} \;
\end{eqnarray*}
(see, {\em e.g.,} \cite{RS75,KR94}), where $ {\bf R} $ and $ {\bf
K} $ correspond to the large scales, and $ {\bf r} $ and $ {\bf k}
$ to the small scales, {\em i.e.,} $ {\bf R} = ({\bf x} +  {\bf
y}) / 2  ,$ $ \quad {\bf r} = {\bf x} - {\bf y},$ $ \quad {\bf K}
= {\bf k}_1 + {\bf k}_2,$ $ \quad {\bf k} = ({\bf k}_1 - {\bf
k}_2) / 2 .$ We assume that there exists a separation of scales,
i.e., the maximum scale of turbulent motions $ l_0 $ is much
smaller then the characteristic scales of inhomogeneities of the
mean fields.

Now we calculate
\begin{eqnarray}
{\partial f_{ij}({\bf k}_1,{\bf k}_2) \over \partial t} &\equiv&
\langle P_{in}({\bf k}_1) {\partial u_{n}({\bf k}_1) \over
\partial t} u_j({\bf k}_2) \rangle
\nonumber\\
&& + \langle u_i({\bf k}_1) P_{jn}({\bf k}_2) {\partial u_{n}({\bf
k}_2) \over \partial t} \rangle \;,
\label{A8}
\end{eqnarray}
where we multiplied equation of motion (\ref{B5}) rewritten in $
{\bf k} $-space by $ P_{ij}({\bf k}) = \delta_{ij} - k_{ij} $ in
order to exclude the pressure term from the equation of motion, $
\delta_{ij} $ is the Kronecker tensor and $ k_{ij} = k_i  k_j /
k^2 .$ Thus, the equations for $ f_{ij}({\bf k, R}) $ is given by
\begin{eqnarray}
{\partial f_{ij} \over \partial t} &=& \hat I_{ijmn}(\bar{\bf U})
f_{mn} + F_{ij} + f_{ij}^{(N)} \;,
\label{A6}
\end{eqnarray}
where
\begin{eqnarray}
\hat I_{ijmn}(\bar{\bf U}) &=& \biggl(2 k_{iq} \delta_{mp}
\delta_{jn} + 2 k_{jq} \delta_{im} \delta_{pn} - \delta_{im}
\delta_{jq} \delta_{np}
\nonumber\\
&& - \delta_{iq} \delta_{jn} \delta_{mp} + \delta_{im} \delta_{jn}
k_{q} {\partial \over \partial k_{p}} \biggr) \nabla_{p} \bar
U_{q} \;, \label{A14}
\end{eqnarray}
and $ f_{ij}^{(N)}({\bf k},{\bf R}) $ is the third moment
appearing due to the nonlinear term, $ \bec{\nabla} = \partial /
\partial {\bf R} ,$ $ F_{ij}({\bf k},{\bf R}) = \langle \tilde F_i
({\bf k},{\bf R}) u_j(-{\bf k},{\bf R}) \rangle + \langle u_i({\bf
k},{\bf R}) \tilde F_j(-{\bf k },{\bf R}) \rangle $ and $ {\bf
\tilde F} ({\bf k},{\bf R},t) = - {\bf k} {\bf \times} ({\bf k}
{\bf \times} {\bf F}^{\rm (st)} ({\bf k},{\bf R})) / k^2 .$
Equation~(\ref{A6}) is written in a frame moving with a local
velocity $ \bar {\bf U} $ of the mean flows. In Eqs.~(\ref{A6})
and (\ref{A14}) we neglected small terms which are of the order of
$O(|\nabla^3 \bar{\bf U}|) .$ Note that Eqs.~(\ref{A6}) and
(\ref{A14}) do not contain terms proportional to $O(|\nabla^2
\bar{\bf U}|) .$ To get Eqs.~(\ref{A6}) and (\ref{A14}) we used an
identity derived in Appendix C.

Equation~(\ref{A6}) for the second moment $f_{ij}({\bf k, R})$
contains third moment $f_{ij}^{(N)}({\bf k, R})$ and a problem of
closing the equations for the higher moments arises. Various
approximate methods have been proposed for the solution of
problems of this type (see, {\em e.g.,} \cite{O70,MY75,Mc90}). The
simplest procedure is the $ \tau $-approximation (the Orszag
third-order closure procedure \cite{O70}). For magnetohydrodynamic
turbulence this approximation was used in \cite{PFL76} (see also
\cite{RK01,KRR90,KMR96}). In the simplest variant, it allows us to
express the deviations of the third moment $f_{ij}^{(N)}({\bf k,
R}) - f_{ij}^{(N0)}({\bf k, R})$ in terms of that for the second
moment $f_{ij}({\bf k, R}) - f_{ij}^{(0)}({\bf k, R})$:
\begin{eqnarray}
f_{ij}^{(N)} - f_{ij}^{(N0)} &=& - {f_{ij} - f_{ij}^{(0)} \over
\tau (k)} \;,
\label{A1}
\end{eqnarray}
where the superscript $ {(0)} $ corresponds to the background
turbulence (it is a turbulence with zero gradients of the mean
fluid velocity, $ \nabla_{i} \bar U_{j} = 0) ,$ and $ \tau (k) $
is the correlation time of the turbulent velocity field. Here we
assumed that the time $ \tau(k) $ is independent of gradients of
the mean fluid velocity because in the framework of the mean-field
approach we may only consider a weak shear: $ \tau_0 |\bec{\nabla}
\bar U| \ll 1 ,$ where $ \tau_0 = l_{0} / u_{0} .$

The $ \tau $-approximation  is in general similar to Eddy Damped
Quasi Normal Markovian (EDQNM) approximation. However some
principle difference exists between these two approaches (see
\cite{O70,Mc90}). The EDQNM closures do not relax to equilibrium,
and this procedure does not describe properly the motions in the
equilibrium state in contrast to the $ \tau $-approximation.
Within the EDQNM theory, there is no dynamically determined
relaxation time, and no slightly perturbed steady state can be
approached \cite{O70}. In the $ \tau $-approximation, the
relaxation time for small departures from equilibrium is
determined by the random motions in the  equilibrium state, but
not by the departure from equilibrium \cite{O70}. As follows from
the analysis by \cite{O70} the $ \tau $-approximation describes
the relaxation to equilibrium state (the background turbulence)
much more accurately than the EDQNM approach.

Note that we applied the $ \tau $-approximation (\ref{A1}) only to
study the deviations from the background turbulence which are
caused by the spatial derivatives of the mean velocity. The
background turbulence is assumed to be known. Here we use the
following model of the background nonhelical, isotropic and weakly
inhomogeneous turbulence:
\begin{eqnarray}
f^{(0)}_{ij}({\bf k},{\bf R}) &=& \frac{1}{8 \pi k^{2}} \biggl(
P_{ij}({\bf k}) + \frac{i}{2 k^2}(k_i \nabla_j
\nonumber\\
&& - k_j \nabla_i) \biggr) u_0^2 \, E(k,{\bf R})  \;, \label{A2}
\end{eqnarray}
where $ \tau(k) = 2 \tau_0 \bar \tau(k) ,$ $ \, E(k) = - d \bar
\tau(k) / dk ,$ $ \, \bar \tau(k) = (k / k_{0})^{1-q} ,$ $ \, 1 <
q < 3 $  is the exponent of the kinetic energy spectrum (e.g., $ q
= 5/3 $ for Kolmogorov spectrum), $ k_{0} = 1 / l_{0} .$

We assume that the characteristic time of variation of the second
moment $f_{ij}({\bf k},{\bf R})$ is substantially larger than the
correlation time $\tau(k)$ for all turbulence scales. Thus in a
steady-state Eq.~(\ref{A6}) reads
\begin{eqnarray}
[\delta_{im} \delta_{jn} &-& \tau \hat I_{ijmn}(\bar{\bf U})]
[f_{mn}({\bf k},{\bf R}) - f_{mn}^{(0)}]
\nonumber\\
&=& \tau \hat I_{ijmn}(\bar{\bf U}) f_{mn}^{(0)} \;,
\label{A3A}
\end{eqnarray}
where we used Eq.~(\ref{A1}). The term $F_{ij}$ in Eq.~(\ref{A6})
determines the background turbulence. The solution of
Eq.~(\ref{A3A}) yields the second moment $f_{ij}({\bf k},{\bf
R})$:
\begin{eqnarray}
f_{ij}({\bf k},{\bf R}) &\approx& f_{ij}^{(0)} + \tau \hat
I_{ijmn}(\bar{\bf U}) f_{mn}^{(0)}({\bf k},{\bf R}) \;, \label{A3}
\end{eqnarray}
where we neglected terms which are of the order of $O(|\tau \nabla
\bar{\bf U}|^2) \ll 1 .$ The first term in Eq.~(\ref{A3}) is
independent of the mean velocity shear and it describes the
background turbulence. The second term in Eq.~(\ref{A3})
determines an effect of the mean velocity shear on turbulence.

\section{Effect of a mean velocity shear on the cross-helicity}

In order to determine the mean electromotive force ${\cal
E}_{i}({\bf r}=0,{\bf R}) \equiv \varepsilon_{inm} \int
g_{mn}({\bf k, R}) \,d {\bf k} ,$ we derive equation for the
cross-helicity tensor:
\begin{eqnarray*}
g_{ij}({\bf k},{\bf R}) &=& \int \langle b_{i}({\bf k} + {\bf
K}/2) u_{j}(-{\bf k} + {\bf K}/2) \rangle
\nonumber\\
&& \times \exp(i {\bf K} {\bf \cdot} {\bf R}) \,d {\bf K} \;,
\end{eqnarray*}
using Eqs.~(\ref{B5}) and (\ref{B6}), i.e., we calculate
\begin{eqnarray}
{\partial g_{ij}({\bf k}_1,{\bf k}_2) \over \partial t} &\equiv&
\langle b_i({\bf k}_1) P_{jn}({\bf k}_2) {\partial u_{n}({\bf
k}_2) \over \partial t} \rangle
\nonumber\\
&& + \langle {\partial b_i({\bf k}_1) \over \partial t} u_j({\bf
k}_2) \rangle \; .
\label{A9}
\end{eqnarray}
This yields equation for $g_{ij}({\bf k, R}) ,$
\begin{eqnarray}
{\partial g_{ij} \over \partial t} = \hat J_{ijmn}(\bar{\bf U})
g_{mn} + \hat M_{ijmn}(\bar{\bf B}) f_{mn} + g_{ij}^{(N)} \;,
\label{A7}
\end{eqnarray}
which describes an effect of a mean velocity shear on the
cross-helicity, where
\begin{eqnarray}
\hat J_{ijmn}(\bar{\bf U}) &=& 2 k_{jq} \delta_{im} \nabla_{n}
\bar U_{q} - \delta_{im} \nabla_{n} \bar U_{j} + \delta_{jn}
\nabla_{m} \bar U_{i}
\nonumber\\
& & + \delta_{im} \delta_{jn} (\nabla_{q} \bar U_{p}) k_{p}
{\partial \over \partial k_{q}} \;,
\label{A15} \\
\hat M_{ijmn}(\bar{\bf B}) &=& \delta_{im} \delta_{jn} \bar B_p
\biggl(i k_p + {1 \over 2} \nabla_{p}\biggr) - \delta_{jn} \bar
B_{i,m}
\nonumber\\
&& - {1 \over 2} \delta_{im} \delta_{jn} \bar B_{p,q} k_{p}
{\partial \over \partial k_{q}} \;, \label{A16}
\end{eqnarray}
and $ g_{ij}^{(N)}({\bf k},{\bf R}) $ is the third moment
appearing due to the nonlinear terms, $ \bar B_{i,j} = \partial
\bar B_{i} / \partial R_{j} .$ Equation~(\ref{A7}) is written in a
frame moving with a local velocity $ \bar {\bf U} $ of the mean
flows. To get Eqs. (\ref{A7})-(\ref{A16}) we used an identity
derived in Appendix C.

Now we use the $ \tau $-approximation which allows us to express
the third moments $ g_{ij}^{(N)}({\bf k},{\bf R}) $ in terms of
the second moments $ g_{ij}({\bf k},{\bf R}) $ :
\begin{eqnarray}
g_{ij}^{(N)} &=& - {g_{ij} \over \tilde \tau (k)} \;,
\label{A4}
\end{eqnarray}
where $ \tilde \tau (k) $ is the characteristic relaxation time,
and we took into account that the cross-helicity tensor $ g_{ij} $
for $\bar{\bf B}=0$ is zero, {\em i.e.,} $g_{ij}(\bar{\bf B}=0) =
0 .$ Note that we applied the $ \tau $-approximation (\ref{A4})
only to study the deviations from the original turbulence (i.e.
the turbulence with $\bar{\bf B}=0)$. These deviations are caused
by a weak mean magnetic field. We considered the case when the
original turbulence does not have magnetic fluctuations. Now we
assume that the characteristic time of variation of the second
moment $g_{ij}({\bf k, R})$ is substantially larger than the
correlation time $ \tau(k) \sim \tilde \tau (k)$ for all
turbulence scales. This allows us to get an equation for the
cross-helicity tensor $g_{ij}({\bf k, R})$ in a steady state:
\begin{eqnarray}
g_{mn}({\bf k, R})[\delta_{im} \delta_{jn} - \tau \hat
J_{ijmn}(\bar{\bf U})] = \tau \hat M_{ijmn}(\bar{\bf B}) f_{mn}
\;, \label{A10}
\end{eqnarray}
where $f_{mn}({\bf k, R})$ in Eq.~(\ref{A10}) is determined by
Eq.~(\ref{A3}). The solution of Eq.~(\ref{A10}) yields
\begin{eqnarray}
g_{ij}({\bf k, R}) &=& \tau \hat M_{ijmn}(\bar{\bf B})
f_{mn}^{(0)} + \tau^2 [\hat M_{ijmn}(\bar{\bf B}) \hat
I_{mnlk}(\bar{\bf U})
\nonumber\\
&& + \hat J_{ijmn}(\bar{\bf U}) \hat M_{mnlk}(\bar{\bf B})]
f_{lk}^{(0)} \;, \label{A11}
\end{eqnarray}
where we neglected terms which are of the order of $O( | \tau
\nabla \bar{\bf U}|^2) \ll 1 .$ The first term in Eq.~(\ref{A11})
is independent of the mean velocity shear and it describes the
isotropic turbulent magnetic diffusion. The second term in
Eq.~(\ref{A11}) determines an effect of the mean velocity shear on
turbulence which causes a modification of the mean electromotive
force, i.e., this term describes an indirect modification of the
mean electromotive force. The last term in Eq.~(\ref{A11})
determines a direct modification of the mean electromotive force
by the mean velocity shear.

\section{The mean electromotive force}

Using Eqs.~(\ref{A3}) and~(\ref{A11}) we determine the mean
electromotive force for an inhomogeneous turbulence with a mean
velocity shear:
\begin{eqnarray}
{\cal E}_{i}({\bf r}=0,{\bf R}) &\equiv& \varepsilon_{inm} \int
g_{mn}({\bf k, R}) \,d {\bf k}
\nonumber\\
&=&  a_{ij} \bar B_{j} + b_{ijk} \bar B_{j,k} \;,
\label{C1}
\end{eqnarray}
where
\begin{eqnarray}
a_{ij} &=& - {l_0^2 \over 18} \biggl\{6 \Lambda_m \biggl[{4q - 9
\over 5} \varepsilon_{imn} (\partial \hat U)_{nj} +
\varepsilon_{ijn} (\partial \hat U)_{mn} \biggr]
\nonumber\\
&& + 3 \varepsilon_{ijm} (\bar{\bf W} {\bf \times}
\bec{\Lambda})_m + \delta_{ij} (\bar{\bf W} \cdot \bec{\Lambda}) -
\bar W_i \Lambda_j \biggr \} \;,
\label{C16}\\
b_{ijk} &=& \beta_{_{T}} \varepsilon_{ijk} - {l_0^2 \over 45}[2
(8q - 11) \varepsilon_{ijm} (\partial \hat U)_{mk}
\nonumber\\
&& + 18 \varepsilon_{ikm} (\partial \hat U)_{mj} + (4q - 17)
\delta_{ij} \bar W_k
\nonumber\\
&& + (4q - 7) \delta_{ik} \bar W_j] \; . \label{C17}
\end{eqnarray}
Here ${\bf \Lambda} = \bec{\nabla} \langle {\bf u}^2 \rangle /
\langle {\bf u}^2 \rangle $ and $l_{0} = \tau_0 u_{0} .$
Equations~(\ref{C16}) and (\ref{C17}) allows us to obtain the
turbulent coefficients defining the mean electromotive force:
\begin{eqnarray}
\alpha_{ij} &=& - {l_0^2 \over 18} \biggl((\bar{\bf W} \cdot
\bec{\Lambda}) \delta_{ij} - {1 \over 2} (\bar W_i \Lambda_j +
\bar W_j \Lambda_i)
\nonumber\\
&& + {3 \over 5} (4 q - 9) \Lambda_m [\varepsilon_{imn} (\partial
\hat U)_{jn} + \varepsilon_{jmn} (\partial \hat U)_{in}] \biggr)
\;,
\nonumber\\
\label{C10}\\
{\bf V}^{\rm (eff)} &=& {l_0^2 \over 30} \biggl((4q + 1) (\partial
\hat U)_{ij} \Lambda_j + {25 \over 6} (\bar{\bf W} {\bf \times}
\bec{\Lambda}) \biggr) \;,
\label{C11}
\end{eqnarray}
and the tensor of turbulent magnetic diffusion $\beta_{ij} ,$ the
$\bec{\delta}$--effect and the tensor $\kappa_{ijk}$ are
determined by Eqs.~(\ref{K3})-(\ref{K5}) with
\begin{eqnarray}
\beta_{0} &=& 2 (5 - 2q) / 45 \;, \quad \delta_0 = 1/9 \;,
\label{C12}\\
\kappa_{1} &=& - 8 (3 - q) / 45 \;, \quad \kappa_{2} = 4 (4q - 1)
/ 45 \;,
\label{C14}
\end{eqnarray}
where we used Eqs.~(\ref{C5})-(\ref{C9}). It is seen from
Eqs.~(\ref{C10}) and (\ref{C11}) that the $\alpha$--effect
described by the tensor $\alpha_{ij}$ and the effective drift
velocity ${\bf V}^{\rm (eff)}$ of the mean magnetic field require
an inhomogeneity of turbulence (i.e., ${\bf \Lambda} \not= 0) .$
The $\kappa$--effect determined by the tensor $\kappa_{ijk}$
arises in an anisotropic turbulence caused by the mean velocity
shear. We will show in the next section that the
$\bec{\delta}$--term in the equation for the mean electromotive
force can cause the mean-field magnetic dynamo in a homogeneous
nonrotating turbulence with an imposed mean velocity shear.

\section{The mean-field magnetic dynamo in a homogeneous
turbulence with a mean shear}

Consider a homogeneous divergence-free turbulence with a mean
velocity shear, e.g., $ \bar{\bf U} = (0, Sx, 0) .$ In this case $
\bar{\bf W} = (0,0,S) ,$ $ \, {\bf \Lambda} = 0 ,$ the
$\alpha$--effect and the effective drift velocity ${\bf V}^{\rm
(eff)}$ of the mean magnetic field vanish. The mean magnetic field
is determined by Eq.~(\ref{E1}), where $\bec{\delta} = \delta_0
l_0^2 \bar{\bf W} $ describes the ''shear-current" effect and
$\beta_{ij} = \beta_{_{T}} \delta_{ij} - 2 \beta_0 l_0^2 (\partial
\hat U)_{ij} $ corresponds to the turbulent magnetic diffusion
with an anisotropic part $ \propto \beta_0 .$ The tensor
$\kappa_{ijk} = l_0^2 \, (\kappa_{1} \, \delta_{ij} \, \bar W_k +
\kappa_{2} \, \varepsilon_{ijm} \, (\partial \hat U)_{mk})$
describes a contribution to the electromotive force related with
the symmetric parts of the gradient tensor of the mean magnetic
field and arises in an anisotropic turbulence caused by the mean
velocity shear. Since the tensor $\kappa_{ijk}$ is multiplied by
the symmetric tensor $(\partial \hat B)_{jk} $ in the the mean
electromotive force, this allows us to rewrite the tensor
$\kappa_{ijk}$ in a more simple but not in a symmetric form. For
simplicity we use the mean magnetic field in the form $ \bar{\bf
B} = (\bar B_x(z), \bar B_y(z), 0) .$ Then Eq.~(\ref{E1}) reduces
to Eqs.~(\ref{E2}) and~(\ref{E3}), where the parameters $\delta_0
,$ $\, \beta_0 ,$ $\, \kappa_1$ and $ \kappa_2$ are determined by
Eqs.~(\ref{C12}) and~(\ref{C14}).

A solution of Eqs.~(\ref{E2}) and~(\ref{E3}) we seek for in the
form $ \propto \exp(\gamma t + i K z) .$ Thus the growth rate of
the mean magnetic field is given by
\begin{eqnarray}
\gamma = S \, l_0 \, K \, \sqrt{\delta_0 - \beta_0 - \kappa_0} -
\beta_{_{T}} \, K^2 \;,
\label{E4}
\end{eqnarray}
where $\kappa_0 = (2 \kappa_1 + \kappa_2) / 4 = (8q - 13) / 45 .$
It follows from Eq.~(\ref{E4}) that the ''shear-current" effect $
(\propto \delta_0) $ causes the generation of the mean magnetic
field, whereas the anisotropic $ (\propto \beta_0) $ and isotropic
$ (\propto \beta_{_{T}}) $ turbulent magnetic diffusions and the
$\kappa$--effect $ (\propto \kappa_0) $ reduce the growth rate of
the mean magnetic field. Note that the maximum growth rate of the
mean magnetic field, $ \gamma_{\rm max} = S^2 l_0^2 (\delta_0 -
\beta_0 - \kappa_0) / 4 \beta_{_{T}} ,$ is attained at $ K = K_m =
S l_0 \sqrt{\delta_0 - \beta_0 - \kappa_0} / 2 \beta_{_{T}} .$
Using expressions for $\delta_0 ,$ $ \, \beta_0$ and $\kappa_0$ we
rewrite the growth rate of the mean magnetic field in the form
\begin{eqnarray}
\gamma = {2 \over 3} \biggl({2 - q \over 5}\biggr)^{1/2} \, S \,
l_0 \, K - \beta_{_{T}} \, K^2 \; . \label{E5}
\end{eqnarray}
Therefore, the generation of the mean magnetic field is possible
when the exponent $ q $ of the energy spectrum of the background
homogeneous turbulence (without imposed mean velocity shear) is
less than 2. Thus, in the Kolmogorov background turbulence with $
q=5/3 $ the mean magnetic field can be generated due to the
''shear-current" effect. The sufficient condition $ \gamma > 0 $
for the dynamo instability reads $ L_B / l_0 > \pi \sqrt{5} /
(\tau_0 S \sqrt{2-q}) ,$ where $ L_B \equiv 2 \pi / K .$

The magnetic dynamo instability due to the ''shear-current" effect
is different from the magnetic instability suggested in
\cite{U02}. The latter instability is caused by the ''negative
turbulent magnetic diffusivity" and is determined by Eq.~(4.2) for
$\bar B_x$ in \cite{U02}. This equation is decoupled from that for
the field $ \bar B_y ,$ i.e., there is no real coupling between
the components of the mean magnetic field  $ \bar B_x $ and $ \bar
B_y .$ In contrast to this, the magnetic dynamo instability due to
the ''shear-current" effect is determined by a system of
equations~(\ref{E2}) and~(\ref{E3}) for the components $ \bar B_x
$ and $ \bar B_y .$ This implies that there is a coupling between
these components of the mean magnetic field. In particular, the
field $ \bar B_y $ generates the field $ \bar B_x $ due to the
$\delta$-term (''shear-current" effect), see the first term in
Eq.~(\ref{E2}). This is similar to the $\alpha$ effect. On the
other hand, the field $ \bar B_x $ generates the field $ \bar B_y
$ due to the pure shear effect (see the first term in
Eq.~(\ref{E3})), like the differential rotation in $\alpha
\Omega$-dynamo. In this sense the instability due to the
''shear-current" effect is a pure magnetic dynamo instability.

However, the above mechanism is different from that for $\alpha
\Omega$-dynamo. Indeed, the dynamo mechanism due to the
''shear-current" effect acts even in homogeneous small-scale
turbulence, while the alpha effect vanishes for homogeneous
turbulence. The difference between these magnetic dynamo
mechanisms can be seen in the form of the growth rate of the mean
magnetic field. Indeed, the generation of the mean magnetic field
is caused by a coupling between the ''shear-current" effect
(described by the first term in Eq.~(\ref{E2}), which is
proportional to the second-order spatial derivative of the mean
magnetic field) and the pure shear effect (described by the first
term in Eq.~(\ref{E3}) which is proportional to the mean magnetic
field). Then the first term in the expression for growth rate in
Eq.~(\ref{E4}) (which is responsible for the generation of the
mean magnetic field due to the ''shear-current" effect) is
proportional to the wave number $K$.

On the other hand, the $\alpha \Omega$-dynamo is caused by a
coupling of the $\alpha$-effect (the corresponding term in the
mean-field equation is proportional to the first-order spatial
derivative of the mean magnetic field) and the differential
rotation (the corresponding term is proportional to the mean
magnetic field). Then the term in the expression for growth rate
of the instability (which is responsible for the generation of the
mean magnetic field in the $\alpha \Omega$-dynamo) is proportional
to $K^{1/2}$.

Note that the properties of the magnetic dynamo caused by the
''shear-current" effect are also different from that for the $
{\bf \Omega} {\times} {\bf J} $--effect. In particular, the mean
magnetic field can be generated due to the $ {\bf \Omega} {\times}
{\bf J} $--effect for an arbitrary exponent $ q $ of the energy
spectrum of the background homogeneous turbulence (see
\cite{RKR02}). The $ {\bf \Omega} {\times} {\bf J} $--effect is
caused by the term $ \bec{\delta} {\bf \times} (\bec{\nabla} {\bf
\times} \bar{\bf B}) $ in the mean electromotive force. In a slow
rotating $ (\Omega \tau_0 \ll 1)$ and homogeneous turbulence $
\bec{\delta} = - (2/9) l_0^2 {\bf \Omega} $ (for details, see
\cite{RKR02}). Note also that the $ {\bf \Omega} {\times} {\bf J}
$--effect cannot generate the mean magnetic field without a
differential rotation.

\section{Conclusions}

In the present paper we discussed a new mechanism of a generation
of a mean magnetic field by a nonrotationg and nonhelical
homogeneous turbulence with an imposed mean velocity shear. This
mechanism is associated with a "shear-current" effect. We showed
that when the exponent of the energy spectrum of the background
turbulence (without the mean velocity shear) is less than 2, a
mean magnetic field can be generated. We calculated the mean
electromotive force for an arbitrary weakly inhomogeneous
turbulence $ (\Lambda l_0 \ll 1)$ with an imposed mean velocity
shear. Inhomogeneity of turbulence and mean velocity shear cause
the $\alpha$--effect and the effective drift velocity of the mean
magnetic field. The ''shear-current" effect was studied using two
different methods: the $\tau$--approximation (the Orszag
third-order closure procedure) and the stochastic calculus (the
path integral representation of the solution of the induction
equation, Feynman-Kac formula and Cameron-Martin-Girsanov theorem,
see Appendixes A and B).

The obtained results may be important in astrophysics, e.g., in
extragalactic clusters and in interstellar clouds. The
extragalactic clusters are nonrotating objects which have a
homogeneous turbulence in the center of a extragalactic cluster.
Sheared motions are created between interacting clusters. The
observed magnetic fields cannot be explained by a small-scale
turbulent magnetic dynamo (see, e.g., \cite{RSS88}). It is
plausible to suggest that the "shear-current" effect can produce a
mean magnetic field in the extragalactic clusters. The sheared
motions can be also formed between interacting interstellar
clouds. The latter can result in a generation of a mean magnetic
field.

\begin{acknowledgments}
We have benefited from useful discussions with A.~Brandenburg and
K.--H.~R\"{a}dler. We are grateful to the two anonymous referees
for their helpful comments and stimulating criticism. This work
was partially supported by INTAS Program Foundation (Grant No.
99-348).
\end{acknowledgments}

\appendix

\section{Investigation of the ''shear-current" effect using stochastic
calculus}

In this Appendix we study the ''shear-current" effect using
stochastic calculus for a random velocity field with a finite
correlation time. In order to derive an equation for the mean
magnetic field we use an exact solution of the induction
equation~(\ref{B2}) in the form of a functional integral for an
arbitrary velocity field taking into account a small yet finite
magnetic diffusion caused by the electrical conductivity of fluid.
This magnetic diffusion, $ D_m ,$ can be described by a random
Brownian motions of a particle. The functional integral implies an
averaging over a random Brownian motions of a particle. The form
of the exact solution used in the present paper allows us to
separate the averaging over both, a random Brownian motions of a
particle and a random velocity field. This method yields the
solution of the induction equation~(\ref{B2}) with an initial
condition $ {\bf h}(t=s,{\bf x}) = {\bf h}(s,{\bf x}) $ in the
form
\begin{eqnarray}
h_{i}(t, {\bf x}) = M_{\bec{\xi}} \{G_{ij}(t,s,\bec{\xi}) \,
\exp(\bec{\xi}^{\ast} \cdot \bec{\nabla}) h_{j}(s, {\bf x}) \} \;
\label{A5}
\end{eqnarray}
(see Appendix B and \cite{KRS02}), where $ \bec{\xi}^{\ast} =
\bec{\xi} - {\bf x} ,$ $ \, G_{ij}(t,s, \bec{\xi}) $ is determined
by equation $ d G_{ij}(t,s,\bec{\xi}) / ds = N_{ik}
G_{kj}(t,s,\bec{\xi}) $ with the initial condition $ G_{ij}(t=s) =
\delta_{ij} .$ Here $ N_{ij} = \partial v_{i} / \partial x_{j} ,$
$ \, M_{\bec{\xi}} \{ \cdot \} $ denotes the mathematical
expectation over the Wiener paths $ \bec{\xi} = {\bf x} -
\int_{0}^{t-s} {\bf v}(t-\sigma,\bec{\xi}) \,d\sigma + (2
D_{m})^{1/2} {\bf w}(t-s) ,$ and the magnetic diffusion, $ D_{m}
,$ is described by a Wiener process $ {\bf w}(t) .$

Consider a random velocity field with a finite constant renewal
time. Assume that in the intervals $ \ldots (- \tau, 0]; (0,
\tau]; (\tau, 2 \tau]; \ldots $ the velocity fields are
statistically independent and have the same statistics. This
implies that the velocity field looses memory at the prescribed
instants $ t = k \tau ,$ where $ k = 0, \pm 1, \pm 2, \ldots .$
This velocity field cannot be considered as a stationary  velocity
field for small times $ \sim \tau ,$ however, it behaves like a
stationary field for $ t \gg \tau .$

In Eq.~(\ref{A5}) we specify instants $ t = (m + 1) \tau $ and $ s
= m \tau .$ Note that the fields $ h_{j}(m \tau, {\bf x}) $ and $
G_{ij}((m + 1) \tau,m \tau,\bec{\xi}) $ are statistically
independent because the field $ h_{j}(m \tau, {\bf x}) $ is
determined in the time interval $ (- \infty, m \tau] ,$ whereas
the function $ G_{ij}((m + 1) \tau,m \tau,\bec{\xi}) $ is defined
on the interval $ (m \tau, (m + 1) \tau] .$ Due to a renewal, the
velocity field as well as its functionals $ h_{j}(m \tau, {\bf x})
$ and $ G_{ij}((m + 1) \tau,m \tau,\bec{\xi}) $ in these two time
intervals are statistically independent. Averaging Eq.~(\ref{A5})
over the random velocity field yields the equation for the mean
magnetic field
\begin{eqnarray}
\bar B_{i}((m + 1) \tau, {\bf x}) &=& M_{\bec{\xi}} \{\langle
G_{ij}(t,s,\bec{\xi}) \, \exp(\bec{\xi}^{\ast} \cdot \bec{\nabla})
\rangle \}
\nonumber\\
&& \times \bar B_{j}(m \tau, {\bf x}) \;,
\label{D6}
\end{eqnarray}
where the operator $ \exp(\bec{\xi}^{\ast} \cdot \bec{\nabla}) $
is determined by
\begin{eqnarray}
\exp(\bec{\xi}^{\ast} \cdot \bec{\nabla}) &=& 1 + \bec{\xi}^{\ast}
\cdot \bec{\nabla} + {1 \over 2!} (\bec{\xi}^{\ast} \cdot
\bec{\nabla})^{2} + \ldots
\nonumber\\
&& + {1 \over m!} (\bec{\xi}^{\ast} \cdot \bec{\nabla})^{m} +
\ldots \;,
\label{D7}
\end{eqnarray}
and the angular brackets $ \langle \cdot \rangle $ denote the
ensemble average over the random velocity field. Note that $
|(\bec{\xi}^{\ast} \cdot \bec{\nabla}) \bar{\bf B}| / |\bar{\bf
B}| \sim l_0 / L_B \ll 1 .$ Thus in the framework of the
mean-field approach we can neglect in Eqs.~(\ref{D6})
and~(\ref{D7}) the terms $\sim O[(\bec{\xi}^{\ast} \cdot
\bec{\nabla})^3 \bar{\bf B}] .$ Now we use the identity
\begin{eqnarray}
\bar B_{i}(t+\tau,{\bf x}) = \exp\biggl(\tau {\partial \over
\partial t} \biggr) \bar B_{i}(t,{\bf x}) \;,
\label{D8}
\end{eqnarray}
which follows from the Taylor expansion
\begin{eqnarray*}
f(t + \tau) = \sum_{m=1}^{\infty} \biggl(\tau {\partial \over
\partial t} \biggr)^{m} f(t) = \exp \biggl(\tau {\partial \over
\partial t} \biggr) {f(t) \over m!} \; .
\end{eqnarray*}
Therefore, Eqs. (\ref{D6})-(\ref{D8}) yield
\begin{eqnarray}
\exp\biggl(\tau {\partial \over \partial t} \biggr) \bar
B_{i}(t,{\bf x}) =  (\bar G_{ij} + \bar G_{ij} \bar \xi_{m}
\nabla_m + A_{ijm} \nabla_m
\nonumber\\
+ C_{ijmn} \nabla_m \nabla_n) \bar B_{j} \equiv \exp(\tau \hat L)
\bar{\bf B} \;,
\label{D9}
\end{eqnarray}
where $\bar G_{ij} = M_{\bec{\xi}} \{\langle G_{ij} \rangle \} =
\delta_{ij} + \bar U_{i,j} \tau + O[(\bec{\nabla} \bar U)^2] ,$
$\, \bar \xi_{i} = M_{\bec{\xi}} \{\langle \xi_{i}^\ast \rangle \}
= - \bar U_{i} \tau + O[(\bec{\nabla} \bar U)^2] ,$ $\, \bar
U_{i,j} = \partial \bar U_{i} / \partial x_{j} ,$ $\, A_{ijm} =
M_{\bec{\xi}} \{\langle G_{ij} \xi^{\ast}_m \rangle \} ,$ $ \,
C_{ijmn} = M_{\bec{\xi}} \{\langle G_{ij} \xi^{\ast}_m
\xi^{\ast}_n \rangle \} ,$ and we introduced the operator $\hat L
,$ which allows us to reduce the integral equation~(\ref{D6}) to a
partial differential equation. Indeed, Eq.~(\ref{D9}), which is
rewritten in the form
\begin{eqnarray}
\exp \biggl[\tau \biggl(\hat L - {\partial \over \partial t}
\biggr) \biggr] \bar{\bf B} = \bar{\bf B} \;,
\label{D11}
\end{eqnarray}
reduces to
\begin{eqnarray}
{\partial \bar{\bf B} \over \partial t} = \hat L \bar{\bf B} \; .
\label{D12}
\end{eqnarray}
The Taylor expansion of the function $\exp(\tau \hat L)$ reads
\begin{eqnarray}
\exp(\tau \hat L) = \hat E + \tau \hat L + (\tau \hat L)^2 / 2 +
... \;,
\label{D14}
\end{eqnarray}
where $\hat E $ is the unit operator. Thus, Eqs.~(\ref{D9})
and~(\ref{D14}) yield
\begin{eqnarray}
\hat L &\equiv& L_{ij} = {1 \over \tau} (\bar G_{ij} - \delta_{ij}
+ \bar G_{ij} \bar \xi_{m} \nabla_m + A_{ijm} \nabla_m)
\nonumber\\
&&+ D_{ijmn} \nabla_m \nabla_n + O(\nabla^3) \;, \label{D15}
\end{eqnarray}
where $ D_{ijmn} = (C_{ijmn} - A_{ikm} A_{kjn}) / 2 \tau .$ Now we
consider homogeneous and nonhelical background turbulence, then
$A_{ijk} = 0$ and the equation for the mean magnetic field is
given by
\begin{eqnarray}
{\partial \bar{B}_i \over \partial t} = [\bec{\nabla} \times
(\bar{\bf U} {\bf \times} \bar{\bf B})]_i + D_{ijmn} \nabla_m
\nabla_n \bar{B}_j \;,
\label{D80}
\end{eqnarray}
where
\begin{eqnarray}
D_{ijmn} = {1 \over 2 \tau} M_{\bec{\xi}} \{\langle G_{ij}
\xi^{\ast}_m \xi^{\ast}_n \rangle \} \; .
\label{D81}
\end{eqnarray}

For a turbulent flow with an imposed mean velocity gradient, the
turbulence is anisotropic. Let us determine the tensor $D_{ijmn}$
in this case. Solution of the equation $ dG_{ij} / ds = N_{ik}
G_{kj}$ with the initial condition $ G_{ij}(t=s) = \delta_{ij} $
is given by
\begin{eqnarray}
G_{ij}(t + \tau,t) = \delta_{ij} + \int_{0}^{\tau}
N_{ij}(t_\sigma,\bec{\xi}) \,d \sigma
\nonumber\\
+ \int_{0}^{\tau} N_{ik}(t_s,\bec{\xi}) \,d s \int_{0}^{s}
N_{kj}(t_\sigma,\bec{\xi}) \,d \sigma + ... \;,
\label{D18}
\end{eqnarray}
which was solved by iterations, where $ t_\sigma = t + \tau -
\sigma .$ Since the velocity field is separated into the mean and
fluctuating parts: $ {\bf v} = \bar{\bf U} + {\bf u} ,$ the tensor
$G_{ij}$ can be presented in the form
\begin{eqnarray}
G_{ij} &=& g_{ij} + G^{(L)}_{ij} \;,
\label{D19}\\
g_{ij} &=& \int_{0}^{\tau} {\partial u_{i}(t_\sigma,\bec{\xi})
\over \partial x_{k}} g_{kj}(t_\sigma,\bec{\xi}) \,d \sigma \;,
\label{D20}\\
G^{(L)}_{ij}(\bar{\bf U}) &=& \bar U_{i,k} \int_{0}^{\tau}
g_{kj}(t_\sigma,\bec{\xi}) \,d \sigma
\nonumber\\
&& + \int_{0}^{\tau} {\partial u_{i}(t_\sigma,\bec{\xi}) \over
\partial x_{k}} G^{(L)}_{kj}(\bar{\bf U}) \,d \sigma \; .
\label{D30}
\end{eqnarray}
Using Eqs.~(\ref{D18})-(\ref{D30}) we obtain
\begin{eqnarray}
G_{ij} = g_{ij} + g_{ik} \bar U_{k,p} \int_{0}^{\tau}
g_{pj}(t_\sigma,\bec{\xi}) \,d \sigma + O[(\bec{\nabla} \bar U)^2]
\; .
\nonumber\\
\label{D31}
\end{eqnarray}
For the derivation of Eq.~(\ref{D31}) we used the identity:
\begin{eqnarray}
(\hat E - \hat X)^{-1} = \hat E + \hat X + \hat X \hat X + \hat X
\hat X \hat X + ... + \;,
\label{D32}
\end{eqnarray}
where $\hat X$ is an arbitrary operator. Similarly, the trajectory
$\xi_{i}^\ast$ can be written in the form:
\begin{eqnarray}
\xi_{i}^\ast = \tilde \xi_{i} - \bar U_{i,k} \int_{0}^{\tau}
\tilde \xi_{k}(t_\sigma,\bec{\xi}) \,d \sigma + O[(\bec{\nabla}
\bar U)^2] \;,
\label{D33}
\end{eqnarray}
where $ \tilde \xi_{i} = - \int_{0}^{\tau} u_{i}(t_\sigma,
\bec{\xi}) \,d \sigma + (2 D_{m})^{1/2} w_i(\tau) .$ Using
Eqs.~(\ref{D31}) and~(\ref{D33}) we determine the tensor $
D_{ijmn} $
\begin{eqnarray}
D_{ijmn} &=& [\langle g_{ij} \tilde \xi_m \tilde \xi_n \rangle -
(\bar U_{m,p} \delta_{nk} + \bar U_{n,p} \delta_{mk}) K_{ijkp}
\nonumber\\
&& + F_{ijmn}] / 2 \tau \;, \label{D34}
\end{eqnarray}
where $F_{ijmn} = \bar U_{k,p} \langle g_{ik} \tilde \xi_m \tilde
\xi_n \int_{0}^{\tau} g_{pj} \,d \sigma \rangle $ and $K_{ijmn} =
\langle g_{ij} \tilde \xi_m \int_{0}^{\tau} \tilde \xi_n \,d
\sigma \rangle .$ In this section hereafter the angular brackets
denote the both averaging: the averaging over a random velocity
field and the averaging over the Wiener trajectories. Now we
determine the tensor $\langle g_{ij} \tilde \xi_m \tilde \xi_n
\rangle ,$ which describes turbulent magnetic diffusion. We take
into account that in a homogeneous and nonhelical turbulence
without mean shear $(\bar U_{i,j} = 0) ,$ the mean-field equation
reads
\begin{eqnarray}
{\partial \bar{B}_i \over \partial t} &=& -[\bec{\nabla} \times
(\beta_{_{T}} \, \bec{\nabla} {\bf \times} \bar{\bf B})]_i
\nonumber\\
&=& - \beta_{_{T}} \varepsilon_{imk} \varepsilon_{knj} \nabla_m
\nabla_n \bar{B}_j \; .
\label{D35}
\end{eqnarray}
In this case $b_{ijk} = \beta_{_{T}} \varepsilon_{ijk} .$
Therefore,
\begin{eqnarray}
\langle g_{ij} \tilde \xi_m \tilde \xi_n \rangle &=& - 2 \tau
\beta_{_{T}} \varepsilon_{imk} \varepsilon_{knj}
\nonumber\\
&\equiv& 2 \tau \beta_{_{T}} (\delta_{ij} \delta_{mn} -
\delta_{in}\delta_{mj}) \;,
\label{D36}
\end{eqnarray}
where
\begin{eqnarray}
\beta_{_{T}} = {1 \over 3} \int_{0}^{\infty} \langle
u_p(0,\bec{\xi}) u_p(\sigma,\bec{\xi}) \rangle \,d \sigma \;,
\label{D37}
\end{eqnarray}
and we used an identity
\begin{eqnarray}
\langle \int_{0}^{\tau} u_i(\mu,\bec{\xi}) \,d \mu \int_{0}^{\tau}
u_j(\sigma,\bec{\xi}) \,d \sigma \rangle
\nonumber\\
= 2 \tau \int_{0}^{\infty} \langle u_i(0,\bec{\xi})
u_j(\sigma,\bec{\xi}) \rangle \,d \sigma \; .
\label{D38}
\end{eqnarray}
The integration of Eq.~(\ref{D36}) over $\tau$ yields the tensor
$K_{ijmn}$:
\begin{eqnarray}
K_{ijmn} = \tau^2 \beta_{_{T}} (\delta_{ij} \delta_{mn} -
\delta_{in}\delta_{mj}) \; .
\label{D39}
\end{eqnarray}
Now we construct the tensor $F_{ijmn} .$ The general form of this
tensor reads
\begin{eqnarray}
F_{ijmn} &=& 2 \tau^2 \beta_{_{T}} (C_1 \bar U_{m,j} \delta_{in} +
C_2 \bar U_{j,m} \delta_{in}
\nonumber\\
&& + C_3 \bar U_{i,j} \delta_{mn} + C_4 \bar U_{j,i} \delta_{mn})
\;, \label{D40}
\end{eqnarray}
where we took into account that the tensor $F_{ijmn}$ in
Eq.~(\ref{D80}) is multiplied by a tensor $\nabla_m \nabla_n
\bar{B}_j$ which is symmetric with respect to indexes $(m,n) .$
Since $ \bec{\nabla} \cdot \bar{\bf B} = 0 ,$ the tensor
$F_{ijmn}$ does not contain the terms with the tensors
$\delta_{jm}$ and $\delta_{jn} ,$  and $F_{ijmn}$ satisfies to the
condition $F_{ijmn} \nabla_i \nabla_m \nabla_n \bar{B}_j = 0 .$
The latter equation yields $C_1 = - C_3$ and $ C_2 = - C_4 .$
Equation~(\ref{D40}) does not have the term $\propto
\varepsilon_{ink} \varepsilon_{jpm} (\partial \hat U)_{pk} $
because we considered a nonhelical turbulence. This is the reason
that Eqs.~(\ref{C17}) and~(\ref{D22}) does not contain the term $
\propto \varepsilon_{jkm} (\partial \hat U)_{mi} $ (cf. with the
general form of the tensor $b_{ijk}$ determined by
Eq.~(\ref{K2})). Therefore, Eqs.~(\ref{D34}), (\ref{D36}),
(\ref{D39}) and~(\ref{D40}) yield the tensor $D_{ijmn} :$
\begin{eqnarray}
D_{ijmn} = \beta_{_{T}} \{\delta_{ij} \delta_{mn} + \tau [C_1
(\bar U_{m,j} \delta_{in} - \bar U_{i,j} \delta_{mn})
\nonumber\\
+ C_2 (\bar U_{j,m} \delta_{in} - \bar U_{j,i} \delta_{mn}) -
(\partial \hat U)_{mn} \delta_{ij}] \} \; .
\label{D21}
\end{eqnarray}
Now we use the identities:
\begin{eqnarray*}
(\bar U_{m,j} \delta_{in} - \bar U_{i,j} \delta_{mn}) \nabla_m
\nabla_n \bar{B}_j &=& [\bec{\nabla} \times (\bec{\nabla} {\bf
\times} (\bar{\bf B} \cdot \bec{\nabla}) \bar{\bf U}) ]_i
\\
(\bar U_{j,m} \delta_{in} - \bar U_{j,i} \delta_{mn}) \nabla_m
\nabla_n \bar{B}_j &=& [\bec{\nabla} \times (\bec{\nabla} {\bf
\times} (\bar{B}_j \bec{\nabla} \bar{U}_j))]_i
\\
(\partial \hat U)_{mn} \nabla_m \nabla_n \bar{B}_i &=&
(\bec{\nabla} \times {\bf Q})_i \;,
\end{eqnarray*}
where $ Q_i = \varepsilon_{ijp} (\partial \hat U)_{pn} \nabla_n
\bar{B}_j .$  The latter identity is derived using the following
identity: $ \varepsilon_{ims} \varepsilon_{sjp} (\partial \hat
U)_{pn} = (\partial \hat U)_{mn} \delta_{ij} - (\partial \hat
U)_{in} \delta_{mj} .$ Note that we neglect the second and higher
order spatial derivatives of the mean velocity. We also neglected
the cross-effect terms which describe an interaction between
molecular and turbulent effects. Thus, Eq.~(\ref{D21}) and the
above identities allows to determine the tensor $b_{ijk}$:
\begin{eqnarray}
b_{ijk} &=& \beta_{_{T}} \varepsilon_{ijk} + \beta_{_{T}} \tau[
(C_1 + C_2) \varepsilon_{ikm} (\partial \hat U)_{mj}
\nonumber\\
&& + {1 \over 2} (C_2 - C_1) \delta_{ij} \bar W_k -
\varepsilon_{ijm} (\partial \hat U)_{mk}] \; .
\label{D22}
\end{eqnarray}
Using Eqs.~(\ref{C5})-(\ref{C9}) and~(\ref{D22}) we determine the
turbulent coefficients defining the mean electromotive force. They
are given by Eqs.~(\ref{K3})-(\ref{K5}) with
\begin{eqnarray}
\beta_{0} = - {1 \over 12} (C_1 + C_2 + 1), \; \; \delta_0 = {1
\over 12} (C_2 - C_1),
\label{D23}\\
\kappa_{1} = - {1 \over 6} (C_2 - C_1), \; \; \kappa_{2} = - {1
\over 3} (C_1 + C_2 - 1), \label{D24}
\end{eqnarray}
where $ l_0 = \sqrt{3 \beta_{_{T}} \tau} .$ For simplicity we use
the mean magnetic field in the form $ \bar{\bf B} = (\bar B_x(z),
\bar B_y(z), 0) .$ Then Eq.~(\ref{E1}) reduces to Eqs.~(\ref{E2})
and~(\ref{E3}), where the parameters $\delta_0 ,$ $\, \beta_0 ,$
$\, \kappa_1$ and $ \kappa_2$ are determined by Eqs.~(\ref{D23})
and~(\ref{D24}). A solution of Eqs.~(\ref{E2}) and~(\ref{E3}) we
seek for in the form $ \propto \exp(\gamma t + i K z) .$ Thus the
growth rate of the mean magnetic field due to the shear-current
effect is given by
\begin{eqnarray}
\gamma \approx S l_0 K \sqrt{C_2 / 3} - \beta_{_{T}} K^2 \;,
\label{D26}
\end{eqnarray}
where we used that $ \sigma_0 = C_2 / 3 .$ Therefore, the magnetic
dynamo instability can be excited when $ C_2 > 0 .$

This approach does not allow us to take into account the effect of
mean velocity shear on turbulence. The method used in this
Appendix only describes the effect of shear on the cross-helicity
tensor. This is one of the reasons that the results obtained by
this method are quantitatively different from that obtained by the
$\tau$ approximation. However, the form of the electromotive force
and a possibility for the large-scale magnetic dynamo in a
homogeneous turbulence due to the shear-current effect are clearly
demonstrated by two different approaches.

\section{Derivation of Eq.~(\ref{A5})}

In order to derive Eq.~(\ref{A5}) we use an exact solution of Eq.
(\ref{B2}) with an initial condition $ {\bf h}(t=s,{\bf x}) = {\bf
h}(s,{\bf x}) $ in the form of the Feynman-Kac formula:
\begin{eqnarray}
h_{i}(t,{\bf x})  = M_{\bec{\xi}} \{G_{ij}(t,s,\bec{\xi}(t,s)) \,
h_{j}(s,\bec{\xi}(t,s))\} \;, \label{T5}
\end{eqnarray}
where the Wiener paths $ \bec{\xi}(t,s) = {\bf x} - \int_{0}^{t-s}
{\bf v}[t-\sigma,\bec{\xi}(t,\sigma)] \,d\sigma + (2 D_{m})^{1/2}
{\bf w}(t-s) .$ Now we assume that
\begin{eqnarray}
h_{i}(t, \bec{\xi}) = \int \exp(i \bec{\xi} \cdot {\bf q})
h_{i}(s, {\bf q}) \,d{\bf q} \; . \label{CC8}
\end{eqnarray}
Substituting Eq.~(\ref{CC8}) into Eq.~(\ref{T5}) we obtain
\begin{eqnarray}
h_{i}(s, {\bf x}) &=& \int M_{\bec{\xi}}
\{G_{ij}(t,s,\bec{\xi}(t,s)) \, \exp[i \bec{\xi}^{\ast} \cdot {\bf
q}] \, h_{j}(s, {\bf q}) \}
\nonumber\\
& & \times \exp(i {\bf q} \cdot {\bf x}) \,d{\bf q} \; .
\label{TC8}
\end{eqnarray}
In Eq.~(\ref{TC8}) we expand the function $ \exp[i
\bec{\xi}^{\ast} \cdot {\bf q}] $ in Taylor series at $ {\bf q} =
0 ,$ i.e., $ \exp[i \bec{\xi}^{\ast} \cdot {\bf q}] =
\sum_{k=0}^{\infty} (1/k!) (i \bec{\xi}^{\ast} \cdot {\bf q})^{k}
.$ Using the identity $ (i {\bf q})^{k} \exp[i {\bf x} \cdot {\bf
q}] = \bec{\nabla}^{k} \exp[i {\bf x} \cdot {\bf q}] $ and
Eq.~(\ref{TC8}) we get
\begin{eqnarray}
h_{i}(t, {\bf x}) &=& M_{\bec{\xi}} \{G_{ij}(t,s,\bec{\xi})
[\sum_{k=0}^{\infty} (1/k!) (\bec{\xi}^{\ast} \cdot
\bec{\nabla})^{k}]
\nonumber\\
& & \times \int h_{j}(s, {\bf q}) \exp(i {\bf q} \cdot {\bf x})
\,d{\bf q} \} \; . \label{BC8}
\end{eqnarray}
After the inverse Fourier transformation in Eq.~(\ref{BC8}) we
obtain Eq.~(\ref{A5}). Equation~(\ref{CC8}) can be formally
considered as an inverse Fourier transformation of the function $
h_{i}(t, \bec{\xi}) .$ However, $ \bec{\xi} $ is the Wiener path
which is not a usual spatial variable. Therefore, it is desirable
to derive Eq.~(\ref{A5}) by a more rigorous method as it is done
below.

To this end we use an exact solution of the Cauchy problem for
Eq.~(\ref{B2}) with an initial condition $ {\bf h}(t=s,{\bf x}) =
{\bf h}(s,{\bf x}) $ in the form
\begin{eqnarray}
h_{i}(t,{\bf x}) = M_{\bec{\zeta}} \{J(t,s,\bec{\zeta}) \tilde
G_{ij}(t,s,\bec{\zeta}) \, h_{j}(s,\bec{\zeta}(t,s)) \} \;,
\label{T2}
\end{eqnarray}
where the matrix $ \tilde G_{ij} $ is determined by the equation $
d \tilde G_{ij}(t,s,\bec{\zeta}) / d s = N_{ik} \tilde
G_{kj}(t,s,\bec{\zeta}) $ with the initial condition $ \tilde
G_{ij}(t=s) = \delta_{ij} ,$ and the function $ J(t,s,\bec{\zeta})
$ is given by
\begin{eqnarray}
J(t,s,\bec{\zeta}) &=& \exp \biggl(- {1\over \sqrt{2 D_{m}}}
\int_{0}^{t-s} v_i(t-\eta,\bec{\zeta}(t,\eta)) \,d w_i(\eta)
\nonumber \\
&&- {1\over 4 D_{m}} \int_{0}^{t-s} {\bf
v}^{2}(t-\eta,\bec{\zeta}(t,\eta)) \,d{\eta} \biggr) \;,
\label{T4}
\end{eqnarray}
$ {\bf w}(t) $ is a Wiener process, and $ M_{\bec{\zeta}} \{ \cdot
\} $ denotes the mathematical expectation over the paths $
\bec{\zeta}(t,s) = {\bf x} + (2 D_{m})^{1/2} ({\bf w}(t) - {\bf
w}(s)) .$ The solution~(\ref{T2}) was first found in \cite{DM84}.
The first integral $ \int_{0}^{t-s} {\bf
v}(t-\eta,\bec{\zeta}(t,\eta)) \cdot \,d{\bf w}(\eta) $ in
Eq.~(\ref{T4}) is the Ito stochastic integral (see, e.g.,
\cite{Mc69}). As follows from Cameron-Martin-Girsanov theorem the
transformation from Eq. (\ref{T5}) to Eq. (\ref{T2}) can be
considered as a change of variables $ \bec{\xi} \to \bec{\zeta} $
in the path integral (\ref{T5}) (see, e.g., \cite{F75}).

The difference between the solutions~(\ref{T5}) and~(\ref{T2}) is
as follows. The function $ h_{j}(s,\bec{\xi}(t,s)) $ in
Eq.~(\ref{T5}) explicitly depends on the random velocity field $
{\bf v} $ via the Wiener path $ \bec{\xi} ,$ while the function $
h_{j}(s,\bec{\zeta}(t,s)) $ in Eq.~(\ref{T2}) is independent of
the velocity $ {\bf v} .$ Trajectories in the Feynman-Kac
formula~(\ref{T5}) are determined by both, a random velocity field
and magnetic diffusion. On the other hand, trajectories in
Eq.~(\ref{T2}) are determined only by magnetic diffusion. Due to
the Markovian property of the Wiener process the
solution~(\ref{T2}) can be rewritten in the form
\begin{eqnarray}
h_{i}(t,{\bf x}) &=& E \{S_{ij}(t,s,{\bf x},{\bf X}') \,
h_{j}(s,{\bf X}') \}
\nonumber\\
& = & \int Q_{ij}(t,s,{\bf x},{\bf x}') h_{j}(s,{\bf x}') \,d {\bf
x}' \;, \label{T8}
\end{eqnarray}
where
\begin{eqnarray}
Q_{ij}(t,s,{\bf x},{\bf x}')  &=& [4 \pi D_{m} (t - s)]^{3/2} \exp
\biggl(- {({\bf x}' - {\bf x})^{2} \over 4 D_{m} (t - s) } \biggr)
\nonumber\\
& & \times S_{ij}(t,s,{\bf x},{\bf x}') \;,
\label{T9}
\end{eqnarray}
$ S_{ij}(t,s,{\bf x},{\bf x}') = M_{\bec{\mu}} \{J(t,s,\bec{\mu})
\tilde G_{ij}(t,s,\bec{\mu}) \} $ and $ \, M_{\bec{\mu}} \{ \cdot
\} $ means the path integral taken over the set of trajectories $
\bec{\mu} $ which connect points $ (t,{\bf x}) $ and $ (s,{\bf
x}') .$  The mathematical expectation $ E \{ \cdot \} $ in
Eq.~(\ref{T8}) denotes the averaging over the set of random points
$ {\bf X}' $ which have a Gaussian statistics (see, e.g.,
\cite{S80}). We used here the following property of the averaging
over the Wiener process $ E \{ M_{\bec{\mu}} \{ \cdot \} \} =
M_{\bec{\zeta}} \{ \cdot \} .$ We considered a random velocity
field with a finite renewal time. Due to a renewal, the velocity
field as well as its functionals $ h_{j}(s, {\bf x}') $ and $
Q_{ij}(t, s, {\bf x}, {\bf x}') $ in the two time intervals are
statistically independent. Now we make a change of variables $
({\bf x},{\bf x}') \to ({\bf x},{\bf x}' = {\bf z}+{\bf x}) $ in
Eq.~(\ref{T8}), i.e., $ \tilde Q_{ij}(t,s,{\bf x},{\bf x}') =
\tilde Q_{ij}(t,s,{\bf x},{\bf z} +{\bf x}) = Q_{ij}(t,s,{\bf
x},{\bf z}) .$ The Fourier transformation in Eq.~(\ref{T8}) yields
\begin{eqnarray*}
h_{i}(t, {\bf x}) = & & \int \int Q_{ij}(t,s,{\bf x},{\bf k})
\exp(i {\bf k} \cdot {\bf z}) \,d {\bf k}
\\
& \times & \int h_{j}(s, {\bf q}) \exp[i {\bf q} \cdot ({\bf
z}+{\bf x})] \,d {\bf q} \,d {\bf z} \; .
\end{eqnarray*}
Since $ \delta({\bf k} + {\bf q}) =  (2 \pi)^{-3} \int \exp[i
({\bf k} + {\bf q}) \cdot {\bf z})] \,d {\bf z} ,$ we obtain that
\begin{eqnarray}
h_{i}(t, {\bf x}) &=& (2 \pi)^{3} \int Q_{ij}(t,s,{\bf x},-{\bf
q}) h_{j}(s, {\bf q})
\nonumber\\
& & \times \exp(i {\bf q} \cdot {\bf x}) \,d{\bf q} \; .
\label{T13}
\end{eqnarray}
In Eq.~(\ref{T13}) the function $ Q_{ij}(t,s,{\bf x},-{\bf q}) $
is given by
\begin{eqnarray}
Q_{ij}(t,s,{\bf x},-{\bf q}) &=& (2 \pi)^{-3} \int Q_{ij}(t,s,{\bf
x},{\bf z})
\nonumber\\
& & \times \exp(i {\bf q} \cdot {\bf z}) \,d{\bf z} \; .
\label{TC2}
\end{eqnarray}
Substituting $ \tilde Q_{ij}(t,s,{\bf x},{\bf x}') =
Q_{ij}(t,s,{\bf x},{\bf z}) $ in Eq.~(\ref{T8}) and taking into
account that $ {\bf x}' = {\bf z} + {\bf x} $ we obtain
\begin{eqnarray}
h_{i}(t,{\bf x}) = \int Q_{ij}(t,s,{\bf x},{\bf z})  h_{j}(s,{\bf
z} + {\bf x}) \,d {\bf z} \; . \label{TC3}
\end{eqnarray}
Equation~(\ref{TC2}) can be rewritten in the form
\begin{eqnarray}
(2 \pi)^{3} Q_{ij}(t,s,{\bf x},&-&{\bf q}) \exp(i {\bf q} \cdot
{\bf x}) = \int Q_{ij}(t,s,{\bf x},{\bf z})
\nonumber\\
& & \times \exp[i {\bf q} \cdot ({\bf z} + {\bf x})] \,d{\bf z}
\; .
\label{TC4}
\end{eqnarray}
The right hand sides of Eqs.~(\ref{TC3}) and~(\ref{TC4}) coincide
when $ {\bf h}(s,{\bf z} + {\bf x}) = {\bf e} \, \exp[i {\bf q}
\cdot ({\bf z} + {\bf x})] ,$ where $ {\bf e} $ is a unit vector.
Thus, a particular solution~(\ref{TC3}) of Eq.~(\ref{B2}) with the
initial condition $ {\bf h}(s, {\bf x}') = {\bf e} \, \exp(i {\bf
q} \cdot {\bf x}') $ coincides in form with the
integral~(\ref{TC4}). On the other hand, a solution of
Eq.~(\ref{B2}) is given by Eq.~(\ref{T2}). Substituting the
initial condition $ {\bf h}(s,\bec{\zeta}) = {\bf e} \, \exp(i
{\bf q} \cdot \bec{\zeta}) = {\bf e} \, \exp[i {\bf q} \cdot ({\bf
x} + (2 D_{m})^{1/2} {\bf w})] $ into Eq.~(\ref{T2}) we obtain
\begin{eqnarray}
h_{i}(t,{\bf x}) &=& M_{\bec{\zeta}} \{J(t,s,\bec{\zeta}) \tilde
G_{ij}(t,s,\bec{\zeta}) e_{j} \,
\nonumber\\
& & \times \exp[i {\bf q} \cdot ({\bf x} + (2 D_{m})^{1/2} {\bf
w})] \} \; .
\label{TC5}
\end{eqnarray}
Comparing Eqs. (\ref{TC3})-(\ref{TC5}) we get
\begin{eqnarray}
Q_{ij}(t,s,{\bf x},-{\bf q}) &=& (2 \pi)^{-3} M_{\bec{\zeta}}
\{J(t,s,\bec{\zeta}) \tilde G_{ij}(t,s,\bec{\zeta}) \,
\nonumber\\
& & \times \exp[i (2 D_{m})^{1/2} {\bf q} \cdot {\bf w}] \} \; .
\label{TC6}
\end{eqnarray}
Now we rewrite Eq.~(\ref{TC6}) using Feynman-Kac
formula~(\ref{T5}). The result is given by
\begin{eqnarray}
Q_{ij}(t,s,{\bf x},-{\bf q}) &=& (2 \pi)^{-3} M_{\bec{\xi}}
\{G_{ij}(t,s,\bec{\xi}(t,s)) \,
\nonumber\\
& & \times \exp[i {\bf q} \cdot \bec{\xi}^{\ast}] \} \;,
\label{TC7}
\end{eqnarray}
where $ \bec{\xi}^{\ast} = \bec{\xi} - {\bf x} .$ Substituting
Eq.~(\ref{TC7}) into Eq.~(\ref{T13}) we obtain
\begin{eqnarray}
h_{i}(t, {\bf x}) &=& \int M_{\bec{\xi}} \{G_{ij}(t,s,\bec{\xi})
\, \exp[i {\bf q} \cdot \bec{\xi}^{\ast}] h_{j}(s, {\bf q}) \}
\nonumber\\
& & \times \exp(i {\bf q} \cdot {\bf x}) \,d{\bf q} \; .
\label{C8C}
\end{eqnarray}
The Fourier transformation in Eq.~(\ref{C8C}) yields
Eq.~(\ref{A5}). The above derivation proves that the
assumption~(\ref{CC8}) is correct for a Wiener path $ \bec{\xi} .$

\section{Identity used for derivation of Eqs.~(\ref{A6}) and (\ref{A7})}

For the derivation of Eqs.~(\ref{A6}) and (\ref{A7}) we used the
following identity
\begin{eqnarray}
&& i k_i \int f_{ij}({\bf k} - {1 \over 2}{\bf  Q}, {\bf K} - {\bf
Q}) \bar U_{p}({\bf  Q}) \exp(i {\bf K} {\bf \cdot} {\bf R}) \,d
{\bf  K} \,d {\bf  Q}
\nonumber\\
& & = -{1 \over 2} \bar U_{p} \nabla _i f_{ij} + {1 \over 2}
f_{ij} \nabla _i \bar U_{p} -  {i \over 4} (\nabla _s \bar U_{p})
\biggl(\nabla _i {\partial f_{ij} \over \partial k_s} \biggr)
\nonumber\\
& & +  {i \over 4} \biggl( {\partial f_{ij} \over
\partial k_s} \biggr) (\nabla _s \nabla _i \bar U_{p}) \; .
\label{D1}
\end{eqnarray}
To derive Eq.~(\ref{D1}) we multiply  the equation
$\bec{\nabla}~\cdot~{\bf u} = 0$ [written in ${\bf k}$-space for
$u_i({\bf k}_1 - {\bf Q})]$ by $u_j({\bf k}_2) \bar U_{p}({\bf Q})
\exp(i {\bf K} {\bf \cdot} {\bf R}) ,$ integrate over ${\bf K}$
and ${\bf Q}$, and average over ensemble of velocity fluctuations.
Here ${\bf k}_1 = {\bf k} + {\bf  K} / 2$ and ${\bf k}_2 = -{\bf
k} + {\bf K} / 2 .$ This yields
\begin{eqnarray}
&& \int i \biggl(k_i + {1 \over 2} K_i - Q_i \biggr) \langle
u_i({\bf k} + {1 \over 2}{\bf K} - {\bf Q}) u_j(-{\bf k} + {1
\over 2}{\bf K}) \rangle
\nonumber\\
& & \times \bar U_{p}({\bf  Q}) \exp{(i {\bf K} {\bf \cdot} {\bf
R})} \,d {\bf  K} \,d {\bf Q} = 0  \; . \label{D2}
\end{eqnarray}
Next, we introduce new variables: $ \tilde {\bf k}_{1} = {\bf k} +
{\bf  K} / 2 - {\bf  Q} ,$ $ \tilde {\bf k}_{2} = - {\bf k} + {\bf
K} / 2 $ and $ \tilde {\bf k} = (\tilde {\bf k}_{1} - \tilde {\bf
k}_{2}) / 2 = {\bf k} - {\bf  Q} / 2,$ $ \tilde {\bf K} = \tilde
{\bf k}_{1} + \tilde {\bf k}_{2} = {\bf  K} - {\bf  Q} .$ This
allows us to rewrite Eq.~(\ref{D2}) in the form
\begin{eqnarray}
& & \int i \biggl(k_i + {1 \over 2} K_i - Q_i \biggr) f_{ij}({\bf
k} - {1 \over 2}{\bf Q}, {\bf K} - {\bf Q}) \bar U_{p}({\bf  Q})
\nonumber\\
& & \times \exp{(i {\bf K} {\bf \cdot} {\bf R})} \,d {\bf  K} \,d
{\bf Q} = 0  \; . \label{D3}
\end{eqnarray}
Since $ |{\bf Q}| \ll |{\bf k}| $ we use the Taylor expansion
\begin{eqnarray}
f_{ij}({\bf k} - {\bf Q}/2, {\bf  K} - {\bf  Q}) \simeq
f_{ij}({\bf k},{\bf  K} - {\bf  Q})
\nonumber\\
- \frac{1}{2} {\partial f_{ij}({\bf k},{\bf  K} - {\bf Q}) \over
\partial k_s} Q_s  + O({\bf Q}^2) \; .
\label{D4}
\end{eqnarray}
We also use the following identities:
\begin{eqnarray}
&& [f_{ij}({\bf k},{\bf R}) \bar U_{p}({\bf R})]_{\bf  K} = \int
f_{ij}({\bf k},{\bf  K} - {\bf  Q}) \bar U_{p}({\bf Q}) \,d {\bf
Q} \;,
\nonumber \\
&& \nabla_{p} [f_{ij}({\bf k},{\bf R}) \bar U_{p}({\bf R})] = \int
i K_{p} [f_{ij}({\bf k},{\bf R}) \bar U_{p}({\bf R})]_{\bf  K}
\nonumber\\
&& \times \exp{(i {\bf K} {\bf \cdot} {\bf R})} \,d {\bf  K} \; .
\label{D5}
\end{eqnarray}
Therefore, Eqs. (\ref{D3})-(\ref{D5}) yield Eq.~(\ref{D1}).

\end{document}